\documentclass[preprint,amssymb]{revtex4-1}
\usepackage{graphicx}
\usepackage{epstopdf}
\usepackage{bm}

\begin{document}

\title{Statistical approach to quantum mechanics \\ II: Nonrelativistic spin}
\date{\today}
\author{G. H. Goedecke}
\affiliation{Physics Department, New Mexico State University, Las Cruces, NM 88003}
\email{ggoedeck@nmsu.edu}

\begin{abstract}

In this second paper in a series, we show that the the general statistical approach to nonrelativistic quantum mechanics developed in the first paper yields a representation of quantum spin and magnetic moments based on classical nonrelativistic spinning top models, using Euler angle coordinates. The models allow half-odd-integer spin and predict supraluminal speeds only for electrons and other leptons, which must be treated relativistically. The spin operators in the space-fixed frame satisfy the usual commutation rules, while those in the rotating body-fixed frame satisfy “left-handed” rules. The commutation rules are independent of the structure of the top, so all nonrelativistic rigidly rotating objects must have integer or odd-half-integer spin. Physical boundary conditions restrict all mixed spin states to involve only half-odd-integer or only integer spin eigenstates. For spin 1/2, the theory automatically yields a modified Pauli-Schr\"odinger equation. The Hamiltonian operator in this equation contains a rigid rotator term and a term involving the square of the magmetic field, as well as an interaction term having the usual form in spherically symmetric and some cylindrically symmetric models, valid for any magnetogyric ratio.


\end{abstract}
\maketitle

\section{Introduction}
\label{intro}

In the first paper in this series, hereafter referred to as [I], we showed that any possible statistical description of all possible nonrelativistic classical motions of any system of particles that are immersed in the stochastic zero-point field (SZPF) always yields the multi-coordinate Schr\"odinger equation with its usual boundary conditions and solutions as an essential statistical equation for the system. We derived the “canonical quantization” rule that the Hamiltonian operator must be the classical Hamiltonian in the $N$-dimensional metric configuration space specified by the classical kinetic energy of the system, with the conjugate momentum $N$-vector replaced by $-i\hbar$ times the vector gradient operator in that space. These results imply that a classical analog of quantum spin should exist for all nonrelativistic rigid rotations of a model particle.  

Many authors have considered classical spinning top models and their possible connection to quantum spin and magnetic moment; we refer to a few examples that seem important in regard to this work~\cite{Bopp,Nyborg,Corben,delaPena,Young,BarutBr,Jauregi,BarutZ,BarutBoMa,Arsenovic}. These treatments either assume commutation rules for the Cartesian components of the spin operator the space-fixed system by analogy with those for orbital angular momentum, or they begin with an Euler angle description of rigid rotation, and simply assume that the momenta conjugate to the angles become operators given by $-i\hbar$ times derivatives with respect to to the angles, in analogy with conjugate translational momenta; both assumptions yield the correct spin commutation rules. However, there are two well-known objections to such models: i) Linear speeds involved in any rigidly rotating object with radius and mass as small as those of an electron must far exceed the vacuum speed of light in order that the model have spin angular momentum of order $\hbar$, and ii) If the spin eigenfunctions must be single-valued periodic functions of the azimuthal Euler angles with period $2\pi$, then half-odd-integer spin is not allowed. 

In this paper, we concentrate on developing the statistical description of a nonrelativistically rotating and translating charged rigid body that is immersed in the SZPF and other specified fields; in the text and in appendix~\ref{appA}, we address and resolve the abovementioned objections. In the interest of readability and simplicity, we follow several authors~\cite{Bopp,delaPena,Young} by using a rigid spherically symmetric extended charged particle model to represent a charged magnetic dipole. (In appendix~\ref{appB}, we treat a model extended charged particle having arbitrary structure.) In Sec.~\ref{classical} we define appropriate Euler angles and obtain the nonrelativistic classical Lagrangian, Hamiltonian, motion equations, metric, and affine connections as needed. In Sec.~\ref{spinops} we utilize the results of [I] to obtain the Hamiltonian operator and the general Schr\"odinger equation for the system. We show that the rotational part of the Hamiltonian operator involves the Cartesian components of the spin angular momentum operator in the space-fixed or the body-fixed frame of reference, expressed in terms of derivatives with respect to the Euler angles. While the space-fixed frame components satisfy the usual commutation rules for spin operators, the body-fixed frame components satisfy “left-handed” commutation rules. We provide arguments to justify half-odd-integer spin. We show how to obtain the simultaneous eigenfunctions of the the space-fixed and body-fixed $z$-components and the square of the spin operator. We find that for spin 1/2, this nonrelativistic theory automatically yields a modified Pauli-Schr\"odinger equation involving the Pauli spin matrices. The modified equation contains a rigid rotator term, a term involving the square of the magnetic field, and an interaction Hamiltonian having the usual form but valid for any magnetogyric ratio.  Furthermore, we show that a particle having half-odd-integer spin cannot access integer spin states, and vice-versa. In Sec.~\ref{summadis} we provide a summary and discussion of the results of this work, including a comparison of orbital and spin angular momentum in rotator models, and a prognosis for future work. In appendix~\ref{appA}, we treat a very simple nonspherical model rotator relativistically to illustrate why linear speeds of mass elements involved in rotation cannot be supraluminal under any conditions. In appendix~\ref{appB}, we treat a nonrelativistic model rotator having arbitrary structure, and show that the Euler angle spin angular momentum operators are indeed completely independent of the model particle structure, whereby any overall nonrelativistic object in a rigid rotator eigenmode must have either odd-half-integer or integer spin.  

\section{Classical nonrelativistic rotator}
\label{classical}

\subsection{Euler angles and angular velocity}
\label{euler}

In order to describe nonrelativistic rotations of any rigid body in classical mechanics, one usually chooses a suitable set of three dimensionless coordinates called Euler angles. (Much of what follows in this subsection is treated in textbooks, but we should present it here for readability and in order to connect with our tensor calculus notation in [I].) One conventional choice of Euler angles is the set $\alpha,\beta,\gamma$ used in several textbooks on mathematical methods of physics, {\it e.g.}, Arfken~\cite{Arfken}. Here, we call these coordinates $\alpha^b$, where indices $b,c,d,...$ from the first part of the alphabet range and sum from 1 to 3. With this set of Euler angles, a general rotation of Cartesian coordinates from a space-fixed system with Cartesian unit basis vectors $\hat{\bm{e}}_i$ to a rotating body-fixed system with Cartesian unit basis vectors $\hat{\overline{\bm{e}}}_i$ is obtained by specifying first a right-hand-screw (RHS) rotation about the space-fixed $z$-axis by azimuthal angle $\alpha^1$, then a RHS rotation about the new $y$-axis by polar angle $\alpha^2,\;0\leq\alpha^2\leq\pi$, then finally a RHS rotation about the new $z$-axis by azimuthal angle $\alpha^3$. The first two of these rotations define an instantaneous axis of rotation. In general, the ranges of the azimuthal angles $\alpha^1,\alpha^3$ should be $(-\infty,\infty)$  because the instantaneous axis of rotation and/or the body may just keep on rotating. However, physical objects should look the same modulo $2\pi$ in all azimuthal angles, so {\em observable} functions should be periodic in $\alpha^1$ and $\alpha^3$ with period $2\pi$. These verbal definitions of this set of Euler angles make it clear that the Cartesian basis vectors in the rotating (barred) system and the space-fixed (unbarred) system are related by
\begin{equation}
\label{eq1}
\hat{\overline{\bm{e}}}_i = R^z_{ij}(\alpha^3)R^y_{jk}(\alpha^2)R^z_{kl}(\alpha^1)\hat{\bm{e}}_l = R_{il}(\alpha)\hat{\bm{e}}_l,
\end{equation}
where the matrices $R^z$ and $R^y$ are given by           	
\begin{equation}
\label{eq2}
R^z(\mu)=\left(\begin{array}{ccc}
    \cos{\mu} & \;\sin{\mu} & \;0 \\
    -\sin{\mu} & \;\cos{\mu} & \;0 \\
    0               &         \;0       & \;1
\end{array}\right);\;\;\;
R^y(\mu)=\left(\begin{array}{ccc}
    \cos{\mu} & \;0 & -\sin{\mu} \\
      0               &         \;1       & 0 \\
    \sin{\mu} & \;0 & \cos{\mu} 
\end{array}\right).
\end{equation}
Thus, the complete rotation is specified by the matrix 
$R(\alpha) = R^z(\alpha^3)R^y(\alpha^2)R^z(\alpha^1)$,
which is orthogonal with determinant +1 (as are the constituent matrices). The set of all such 3X3 matrices forms the defining irreducible representation of the rotation group $O^+_3$. 

The angular velocity 3-vector can be found from the relations defining rigidly rotating Cartesian coordinates,
\begin{equation}
\label{eq3}
d\hat{\overline{\bm{e}}}_i/dt = \bm{\omega}\times\hat{\overline{\bm{e}}}_i
\end{equation}
where $\bm{\omega}$ is the instantaneous angular velocity. Its Cartesian components $\omega_i = \hat{\bm{e}}_i\bm{\cdot}\bm{\omega}$ in the space-fixed frame, and $\overline{\omega}_i = \hat{\overline{\bm{e}}}_i\bm{\cdot}\bm{\omega}$  in the rotating frame, can be obtained using eqs.~(\ref{eq1})-(\ref{eq3}). The results are
\begin{equation}
\label{eq4}
\omega_i = a_{ib}\dot{\alpha}^b;\;\;\;\overline{\omega}_i = b_{ib}\dot{\alpha}^b = R_{ik}\omega_k,
\end{equation}
where we have specified the trajectories of the Euler angles by $\alpha^b = \alpha^b(t)$. The matrices $(a)$ and $(b)$ are given by
\begin{equation}
\label{eq5}
(a) = \left(\begin{array}{ccc}
   0     &  \;-\sin{\alpha^1} & \sin{\alpha^2}\cos{\alpha^1} \\
   0     &   \;\;\cos{\alpha^1} & \sin{\alpha^2}\sin{\alpha^1} \\
   1     &             \;0            &                \cos{\alpha^2}
\end{array}\right);\;\;\;
(b) = \left(\begin{array}{ccc}
   -\sin{\alpha^2}\cos{\alpha^3} & \sin{\alpha^3} & 0 \\
   \;\sin{\alpha^2}\sin{\alpha^3} & \cos{\alpha^3} & 0 \\
                \;\cos{\alpha^2}           &          0             & 1
\end{array}\right).
\end{equation}
Note that eq.~(\ref{eq4}) implies $(b) = (R)(a)$. The inverse of the matrix $(a)$ will also be needed:     
\begin{equation}
\label{eq6}
(a)^{-1} = \left(\begin{array}{ccc}
   -\cos{\alpha^1}\cot{\alpha^2} & -\sin{\alpha^1}\cot{\alpha^2} & 1 \\
                  -\sin{\alpha^1}         &                      \cos{\alpha^1}    & 0 \\
   \cos{\alpha^1}/\sin{\alpha^2} & \sin{\alpha^1}/\sin{\alpha^2} & 0
\end{array}\right).
\end{equation}

\subsection{Classical nonrelativistic rotator dynamics}
\label{classicaldynamics}

As mentioned above, we first consider a very simple model particle, a rigid extended spherically symmetric object having the attributes of electric charge $q$, mass $m$, moment of inertia $I$, and the 3-vectors CM position $\bm{X}(t)$, intrinsic magnetic dipole moment $\bm{\mu}(t)$, and angular velocity $\bm{\omega}(t)$, both of the latter about the CM. (In Appendix B, we consider a rigid object of arbitrary shape and structure.) The classical definition of intrinsic magnetic moment (Gaussian units) is
\begin{equation}
\label{eq7}
\bm{\mu}(t) = (2c)^{-1}\int d^3x' \bm{x}'\times\bm{J}(\bm{x},t),
\end{equation}
where $\bm{x}'=\bm{x}-\bm{X}(t)$, and $\bm{J}(\bm{x},t)$ is the electric current density. For a spherically symmetric translating and rigidly rotating model,  
\begin{equation}
\label{eq8}
\bm{J}(\bm{x},t) = q[\dot{\bm{X}}(t) + \bm{\omega}(t)\times\bm{x}']f_q(x'),
\end{equation}
where $x' = |\bm{x}'|$,  and the electric charge density is $qf_q(x')$, so that $\int d^3x' f_q(x') = 1$. Combining these equations and noting that $\int d^3x' \bm{x}' f_q(x') = 0$ yields
\begin{equation}
\label{eq9}
\bm{\mu}(t) = q(2c)^{-1}\left[\mbox{\small$\frac{2}{3}$}\int d^3x' x'^2 f_q(x')\right]\bm{\omega}(t).
\end{equation}
The nonrelativistic kinetic spin angular momentum is
\begin{equation}
\label{eq10}
\bm{S}^K(t) = m\int d^3x' [\bm{x}'\times(\bm{\omega}(t)\times\bm{x}')]f_m(x') = I\bm{\omega}(t),
\end{equation}
where $mf_m(x')$ is the mass density, so that $\int d^3x' f_m(x') = 1$, and the moment of inertia $I$ is given by
\begin{equation}
\label{eq11}
I =\mbox{\small$\frac{2}{3}$} m\!\int d^3x' x'^2 f_m(x').
\end{equation}
Combining eqs.~(\ref{eq9})-(\ref{eq11}) yields
\begin{equation}
\label{eq12}
\bm{\mu}(t) = (gq/2mc)\bm{S}^K = \tilde{g}I\bm{\omega}(t),
\end{equation}
where the dimensionless parameter g is defined by
\begin{equation}
\label{eq13}
g = \int d^3x' x'^2 f_q(x')/\int d^3x' x'^2 f_m(x'),
\end{equation}
and the magnetogyric ratio $\tilde{g}$ is defined by
\begin{equation}
\label{eq14}
\tilde{g} = gq/2mc.
\end{equation}
We provided the detailed derivation above not only for clarity but also to emphasize that the intrinsic magnetic moment of a spherically symmetric rigidly rotating charged body is proportional to the {\it kinetic} spin angular momentum $\bm{S}^K$, not to the spin angular momentum $\bm{S}$ that is canonically conjugate to the Euler angles, to be defined below. This point has been emphasized by several authors~\cite{Young,BarutBoMa,Arsenovic}.

     The dimensionless parameter $g = 2$ for a bare electron, but may have quite different values for other particles. (As we shall see in what follows, the classical description of rotation and/or ``zitterbewegung" for electrons and probably for other leptons must be relativistic).  If the charge density of the particle is proportional to its mass density, then $g = 1$. The conventional nonrelativistic Lagrangian for the charged translating rigid rotator system considered here can be derived quite easily as the sum of the translational kinetic energy (KE) of the CM, the rotational KE about the CM, and the interaction Lagrangian 
$$L_{int} = \int d^3x \left( -\rho_q\varphi + c^{-1}\bm{J}\bm{\cdot}\bm{A}\right)$$
for the interaction of external electromagnetic potential fields $(\varphi,\bm{A})$ with any charge-current densities $(\rho_q,\bm{J})$. For the spherical model particle used here, sharply localized around $\bm{x} = \bm{X}(t)$, the appropriate approximation for the interaction Lagrangian is obtained by a Taylor expansion of the external fields about the CM. The expression for $L$ that is valid through dipole moment interactions is
\begin{equation}
\label{eq15}
L = [\mbox{$\frac{1}{2}$}m\tilde{\bm{V}}^2 - q\varphi + c^{-1}q\bm{A}\bm{\cdot}\tilde{\bm{V}}] + [\mbox{$\frac{1}{2}$}I\bm{\omega}^2 + \tilde{g}I\bm{\omega}\bm{\cdot}\bm{B}],
\end{equation}
where $\tilde{\bm{V}}$ is the 3-vector CM velocity, and the fields $\varphi,\bm{A}$ and the magnetic field (flux density) $\bm{B}$ are evaluated at the CM. Clearly, the first bracket is $L_{tr}$, the Lagrangian involving the translational motion of the CM, and the second bracket is $L_{rot}$, the Lagrangian involving the rotational motion, including the usual interaction $\bm{\mu}\bm{\cdot}\bm{B}$ of a magnetic dipole moment with a magnetic field.  Of course, if the magnetic field is anything other than a constant vector, this interaction term influences the translational motion as well, as will be discussed below. Note that the electric dipole moment about the CM is zero for this model particle, since the CM is also the center of charge.
    
The Hamiltonian can be obtained without specifying the rotational coordinates. First, define the conjugate momentum 3-vectors
\begin{equation}
\label{eq16}
\tilde{\bm{P}} = \partial L/\partial\tilde{\bm{V}} = m\tilde{\bm{V}} + q\bm{A}/c;
\end{equation}
\begin{equation}
\label{eq17}
\bm{S} = \partial L/\partial\bm{\omega} = I\bm{\omega} + I\tilde{g}\bm{B},
\end{equation}
where $\tilde{\bm{P}}$ is the translational momentum conjugate to the CM velocity $\tilde{\bm{V}}$, and $\bm{S}$ is the intrinsic (spin) angular momentum conjugate to $\bm{\omega}$. Then the Hamiltonian is given as usual by $H = \tilde{\bm{V}}\bm{\cdot}\tilde{\bm{P}} + \bm{\omega}\bm{\cdot}\bm{S} - L$. Applying eqs.~(\ref{eq15}) – (\ref{eq17}) yields
\begin{equation}
\label{eq18}
H = \mbox{$\frac{1}{2m}$}(\tilde{\bm{P}}-q\bm{A}/c)^2 + \mbox{$\frac{1}{2I}$}(\bm{S}-I\tilde{g}\bm{B})^2 + q\varphi.
\end{equation}
This Hamiltonian is clearly equal to the sum of the kinetic and potential energies. It is conserved if neither of the potentials $\bm{A}$ and $\varphi$ depend explicitly on the time. Note that the effective “interaction Hamiltonian” involving the spin is the cross-term in the second term, $-\tilde{g}\bm{B}\bm{\cdot}\bm{S}$, which is generally misinterpreted as $-\bm{\mu}\bm{\cdot}\bm{B}$. This form of the classical Hamiltonian for a system of one extended spherically symmetric rotating charged particle in electromagnetic fields was presented {\it e.g.} by Young~\cite{Young}, but has not been included in most standard textbooks, despite the fact that once the angular velocity is expressed in terms of a set of Euler angles, Hamilton's canonical equations yield the correct Euler-Lagrange equations of motion for the rotation only if the second term in eq.~(\ref{eq18}) is present in its entirety. 

Now, using eq.~(\ref{eq4}), we express the rotational kinetic energy $T_{rot}$ in terms of the Euler angles:
\begin{equation}
\label{eq19}
T_{rot} = \mbox{$\frac{1}{2}$}I\omega_i\omega_i = \mbox{$\frac{1}{2}$}m[\mbox{$\frac{I}{m}$}a_{ib}a_{ic}\dot{\alpha}^b\dot{\alpha}^c];
\end{equation} 
\begin{equation}
\label{eq20}
T_{rot} = \mbox{$\frac{1}{2}$}I\bar{\omega}_i\bar{\omega}_i = \mbox{$\frac{1}{2}$}m[\mbox{$\frac{I}{m}$}b_{ib}b_{ic}\dot{\alpha}^b\dot{\alpha}^c].
\end{equation}
These expressions immediately reveal the covariant metric $g^{rot}_{bc} = \mbox{$\frac{I}{m}$}a_{ib}a_{ic} = \mbox{$\frac{I}{m}$}b_{ib}b_{ic}$ in the Euler angle 3-space. Since $g^{rot}_{bc} = \bm{e}_b\bm{\cdot}\bm{e}_c$ in terms of the covariant basis vectors in that space, and $g_{rot}^{bc} = \bm{e}^b\bm{\cdot}\bm{e}^c$, and also $\delta^c_b = \bm{e}_b\bm{\cdot}\bm{e}^c$ , these basis vectors must satisfy 
\begin{equation}
\label{eq21}
\bm{e}_b =\mbox{$\sqrt{\frac{I}{m}}$}a_{ib}\hat{\bm{e}}_i =\mbox{$\sqrt{\frac{I}{m}}$}b_{ib}\hat{\bar{\bm{e}}}_i;
\end{equation}
\begin{equation}
\label{eq22}
\bm{e}^c =\mbox{$\sqrt{\frac{m}{I}}$}a^{-1}_{cj}\hat{\bm{e}}_j =\mbox{$\sqrt{\frac{m}{I}}$}b^{-1}_{cj}\hat{\bar{\bm{e}}}_j;
\end{equation}
These relations are examples of some of the possibilities discussed in the Appendix of [I]. For example, they imply that $g^{cd}_{rot} = \mbox{$\frac{m}{I}$}a^{-1}_{cj}a^{-1}_{dj}$, which in turn implies $g^{rot}_{bc}g_{rot}^{cd} = \delta_b^d$, which must be the case. Also, it is straightforward to show from eqs.~(\ref{eq21}) and (\ref{eq22}) that the affine connections $\Gamma^d_{bc}$, which are defined by eq. (A10) of [I], $\partial_b\bm{e}_c = \Gamma^d_{bc}\bm{e}_d$, are not symmetric under interchange of their lower indices, whereby the Euler angle space is a space with torsion, and the affine connections are not equal to the corresponding Christoffel symbols. But the connections may all be evaluated using eqs.~(\ref{eq21}) and (\ref{eq22}) and their inverses. We will need only one of the connections in this paper, which we derive below. 

     The 3X3 matrix of the covariant components of the metric is given by $g^{rot}_\cdot = \mbox{$\frac{I}{m}$}(a)^T(a) = \mbox{$\frac{I}{m}$}(b)^T(b)$; both expressions yield
\begin{equation}
\label{eq23}
(g^{rot}_\cdot) = \frac{I}{m}\left(
\begin{array}{ccc}
         1                   &0       &\cos{\alpha^2} \\
         0                   &1       &0                  \\
    \cos{\alpha^2}  &0        &1
\end{array}\right)
\end{equation}
Each covariant metric component has dimension (length)$^2$. We may show fairly easily from eqs.~(\ref{eq5}),(\ref{eq6}), and (\ref{eq22}) that
$$\partial_b\bm{e}^b = (\partial_ba^{-1}_{bj})a_{jc}\bm{e}^c = -\Gamma^b_{bc}\bm{e}^c = -\cot{\alpha^2}\bm{e}^2$$
Also, from eq.~(\ref{eq23}), $\sqrt{|g_\cdot|} = (I/m)^{3/2}\sin{\alpha^2}$, so that 
$$\bm{e}^c\partial_c\ln{\sqrt{|g_\cdot|}} = (\cot{\alpha^2})\bm{e}^2$$.
Therefore, the identity (A18) in [I], $\int d^Nx\sqrt{|g_\cdot|}\bm{\nabla}f = 0$, where $f$ is any function of the coordinates $x^1,x^2,x^3,...$ of an $N$-space and the integration extends over all $N$-space, is valid for the 3-dimensional Euler angle subspace here, despite the asymmetry of the affine connections. This result is important in what follows. Writing out $T_{rot}$ yields
\begin{equation}
\label{eq24}
T_{rot} = \mbox{$\frac{1}{2}$}I[(\dot{\alpha}^1)^2 + (\dot{\alpha}^3)^2 + 2\dot{\alpha}^1\dot{\alpha}^3\cos{\alpha^2} + (\dot{\alpha}^2)^2].
\end{equation}
The magnetic interaction term in the Lagrangian is then
\begin{equation}
\label{eq25}
L^{mag}_{int} = \tilde{g}I\bm{\omega}\bm{\cdot}\bm{B} = \tilde{g}IB_ia_{ib}\dot{\alpha}^b = \tilde{g}I\bar{B}_ib_{ib}\dot{\alpha}^b,
\end{equation}
where here the $\bar{B}_i$ are {\em defined} by $\bar{B}_i = R_{ik}B_k$, as if they were the barred (rotating) frame Cartesian components of an ordinary 3-vector. Note that the $\bar{B}_i$ depend on the Euler angles and thus are time-dependent even if the $B_i$ are not. Thus, the rotational part of the Lagrangian~(\ref{eq15}) may be expressed as
\begin{equation}
\label{eq26}
L_{rot} = \mbox{$\frac{1}{2}$}mg^{rot}_{bc}\dot{\alpha}^b\dot{\alpha}^c + \tilde{g}IB_ia_{ib}\dot{\alpha}^b = \mbox{$\frac{1}{2}$}mg^{rot}_{bc}\dot{\alpha}^b\dot{\alpha}^c + \tilde{g}I\bar{B}_ib_{ib}\dot{\alpha}^b.
\end{equation}
The momenta conjugate to the angles are
\begin{equation}
\label{eq27}
P_b = \partial L/\partial\dot{\alpha}^b = Ia_{ib}(a_{ic}\dot{\alpha}^c +\tilde{g}B_i) = Ib_{ib}(b_{ic}\dot{\alpha}^c +\tilde{g}\bar{B}_i).
\end{equation}
These momenta have the dimension of angular momentum. Contraction with $a^{-1}_{bk}$ and $b^{-1}_{bk}$ and comparison with eq.~(\ref{eq17}) yields 
\begin{equation}
\label{eq28}
S_k = a_{bk}^{-1}P_b =I(\omega_k + \tilde{g}B_k);\;\;\;\bar{S}_k = b_{bk}^{-1}P_b = (\bar{\omega}_k +  \tilde{g}\bar{B}_k),
\end{equation}
where the $(S_k, \bar{S}_k)$ are the (space-fixed, body-fixed) Cartesian components of the conjugate spin angular momentum 3-vector. Then, from the definition of the rotational part of the Hamiltonian, $H_{rot} = \dot{\alpha}^bP_b - L_{rot}$, where $L_{rot}$ is given by eq.(\ref{eq26}), one obtains easily
\begin{equation}
\label{eq29}
H_{rot} = \mbox{$\frac{1}{2I}$}(\bm{S} - I\tilde{g}\bm{B})^2,
\end{equation}
which is the rotational part of eq.~(\ref{eq18}). Altogether, the classical Lagrangian and Hamiltonian in terms of Cartesian CM coordinates and the three Euler angle coordinates are given by
\begin{equation}
\label{eq30}
L = [\mbox{$\frac{1}{2}$}m\delta_{ij}\dot{X}^i\dot{X}^j - q\varphi + qc^{-1}\dot{X}^iA_i] + 
[\mbox{$\frac{1}{2}$}mg^{rot}_{bc}\dot{\alpha}^b\dot{\alpha}^c + \tilde{g}IB_ia_{ib}\dot{\alpha}^b],
\end{equation}
\begin{equation}
\label{eq31}
H = \mbox{$\frac{1}{2m}$}(\tilde{P}_i - qc^{-1}A_i)(\tilde{P}_i - qc^{-1}A_i) + q\varphi + \mbox{$\frac{1}{2I}$}(S_i - I\tilde{g}B_i)(S_i - I\tilde{g}B_i),
\end{equation}
where we have used the space-fixed frame components $S_i = a^{-1}_{bi}P_b$ and $B_i$ since the latter are presumed known and may be constants, as mentioned above.

We may rewrite the Lagrangian~(\ref{eq30}) in the general form of eq. (21) of [I] by going to six dimensions and renaming a few quantities. However, the motion equations are more easily derived from eqs.~(\ref{eq30}, \ref{eq31}) directly. After some algebra, the Euler-Lagrange (EL) equations applied to eq.~(\ref{eq30}) yield the following 3-vector classical motion equations for translation and rotation:
\begin{equation}
\label{eq32}
m\dot{\tilde{\bm{V}}} = q(\bm{E} + c^{-1}\tilde{\bm{V}}\times\bm{B}) + \tilde{g}I\bm{\nabla}\bm{B} \bm{\cdot}\bm{\omega},
\end{equation}
\begin{equation}
\label{eq33}
I\dot{\bm{\omega}} = \tilde{g}I\bm{\omega}\times\bm{B} - \tilde{g}I\tilde{\bm{V}}\bm{\cdot}\bm{\nabla B}.
\end{equation}
The last term in eq.~(\ref{eq32}) is the expected force $\bm{\nabla}\bm{B}\bm{\cdot}\bm{\mu}$ on a magnetic dipole moment in any (nonuniform) magnetic field, time-dependent or not. The second term in eq.~(\ref{eq33}) is the expected torque $\bm{\mu}\times\bm{B}$, while the last term is another torque that is omitted from most textbook presentations. As emphasized by Young~\cite{Young}, that torque must present in order to predict conservation of the total kinetic energy $\mbox{$\frac{1}{2}$}m\tilde{\bm{V}}^2 + \mbox{$\frac{1}{2}$}I\bm{\omega}^2$ for the case of a nonuniform static magnetic field and zero electric field. During the course of deriving eqs.~(\ref{eq32}) and (\ref{eq33}) from the EL equations, the following useful identity must be proved:
\begin{equation}
\label{eq34}
(a^{-1}_{bi}a^{-1}_{cj} - a^{-1}_{bj}a^{-1}_{ci})(\partial a_{kc}/\partial \alpha^b) = \epsilon_{ijk},
\end{equation}
where $\epsilon_{ijk}$ is the Levi-Civita completely antisymmetric three-index symbol. This identity was not trivial to prove. (The author could not find a simple general derivation, and resorted to brute force, by calculating and verifying the identity for each symbol, starting from eqs.~(\ref{eq5}) and (\ref{eq6}) for the matrices $(a)$ and $(a)^{-1}$.)  A similar identity exists among the elements of $(b)$. We note that if and only if $\bm{B}$ has no intrinsic time dependence, one may apply eq.~(\ref{eq17}) and write eq.~(\ref{eq33}) as $\dot{\bm{S}} = \tilde{g}\bm{S}\times\bm{B}$, because in such a case $\tilde{\bm{V}}\bm{\cdot}\bm{\nabla}\bm{B} = d\bm{B}/dt$.  

\section{Spin operators and statistical wave equation}
\label{spinops}

\subsection{Wave equation for arbirary spin}
\label{genspin}

The relevant statistical wave equation for any nonrelativistic system having six coordinates is the six-dimensional version of the general statistical wave equation (35) of [I], with Hamiltonian operator given by Eq. (34) of [I]. However, as shown above, for the model system considered here the classical Hamiltonian may be written as eq.~(\ref{eq31}). One need only note that the rotational part of the 6D gradient operator in eqs.\,(31)-(34) of [I] is $\bm{e}^b\partial_b$, so that in eq.~(\ref{eq31}) above the conjugate classical momenta are replaced by the momentum operators
\begin{equation}
\label{eq35}
\tilde{P}_i \rightarrow p^{op}_i = -i\hbar\partial_i;\;\;\;P_b \rightarrow p^{op}_b = -i\hbar\partial_b,
\end{equation}
where here $\partial_b = \partial/\partial\alpha^b$. Making these substitutions in eq.~(\ref{eq31}) yields
\begin{equation}
\label{eq36}
H^{op} = \mbox{$\frac{1}{2m}$}(p^{op}_i - qc^{-1}A_i)(p^{op}_i - qc^{-1}A_i) + q\varphi + \mbox{$\frac{1}{2I}$}(S^{op}_i - I\tilde{g}B_i)(S^{op}_i - I\tilde{g}B_i),
\end{equation}
where, from eqs.~(\ref{eq28}) and (\ref{eq35}),
\begin{equation}
\label{eq37}
S^{op}_i = a^{-1}_{bi}p^{op}_b = -i\hbar a^{-1}_{bi}\partial_b.
\end{equation}
Just after eq.(\ref{eq23}), we showed that eq.\,(A18) of [I] is valid for the Euler angle 3-space, which ensures that the operators $H^{op}$ and $\bm{S}^{op}$ are Hermitian, as discussed in general in [I]. It is easy to show that the spin operators $S^{op}_i$  satisfy the usual commutation rules for angular momentum,
\begin{equation}
\label{eq38}
[S_i^{op},S_j^{op}] = i\hbar\epsilon_{ijk}S_k^{op}.
\end{equation}
One way to obtain this result is to combine eqs.~(\ref{eq37}) and (\ref{eq34}). Another way is to write out eq.~(\ref{eq37}) as a matrix equation, using eqs.~(\ref{eq5}) and (\ref{eq6}), and then calculate each commutator directly. Eqs.~(\ref{eq28}) and (\ref{eq35}) also yield the expressions
\begin{equation}
\label{eq39}
\bar{S}_i^{op} = b^{-1}_{bi}p^{op}_b = -i\hbar b^{-1}_{bi}\partial_b.
\end{equation}
Using this equation, and eq.~(\ref{eq5}) to obtain the matrix $(b)$ and its inverse, it is straightforward to show that the rotating system spin operators $\bar{S}^{op}_i$ satisfy
\begin{equation}
\label{eq40}
[\bar{S}_i^{op},\bar{S}_j^{op}] = -i\hbar\epsilon_{ijk}\bar{S}_k^{op}.
\end{equation}
Note the minus sign, compared to eq.~(\ref{eq38})! To the best of our knowledge, these “left-handed” rotating system commutation relations are not mentioned in quantum mechanics textbooks. One would expect that they have been presented in the literature, but we have been unable to locate a reference. After some algebra, either eq.~(\ref{eq37}) or eq.\,(\ref{eq39}) yields the expression for $(\bm{S}^{op})^2$ in terms of the Euler angles:
\begin{equation}
\label{eq41}
(\bm{S}^{op})^2 = -\hbar^2\left[\partial^2_{\alpha^2} + \cot{\alpha^2}\partial_{\alpha^2} + (\sin{\alpha^2})^{-2}(\partial^2_{\alpha^1} + \partial^2_{\alpha^3} - 2\cos{\alpha^2}\partial_{\alpha^1}\partial_{\alpha^3})\right].
\end{equation}
The operators $(\bm{S}^{op})^2, S^{op}_3, \bar{S}^{op}_3$ all commute and thus have simultaneous eigenfunctions. These eigenfunctions, sometimes called Wigner harmonics, are proportional to the elements of the matrices of the irreducible representations of the group SU(2). We shall use Dirac notation and denote them by $\left|s,m_s,\bar{m}_s\right>$. They satisfy
\begin{equation}
\label{eq42}
(\bm{S}^{op})^2\left|s,m_s,\bar{m}_s\right> = s(s+1)\hbar^2\left|s,m_s,\bar{m}_s\right>,
\end{equation}
\begin{equation}
\label{eq43}
S^{op}_3\left|s,m_s,\bar{m}_s\right> = m_s\hbar\left|s,m_s,\bar{m}_s\right>,\;\;\;\bar{S}^{op}_3\left|s,m_s,\bar{m}_s\right> = \bar{m}_s\hbar\left|s,m_s,\bar{m}_s\right>,
\end{equation}
where $s = 0,1/2,1,3/2,2,...,$ and $m_s$ and $\bar{m}_s$ run independently from $s$ to $-s$ in integer steps. In terms of the Euler angles, all these spin eigenfunctions have the general form~\cite{MathewsW}
\begin{equation}
\label{eq44}
\left|s,m_s,\bar{m}_s\right> = \exp{i(m_s\alpha^1 + \bar{m}_s\alpha^3)}u^s_{m_s,\bar{m}_s}(\alpha^2),
\end{equation}
where the $u^s_{m_s,\bar{m}_s}(\alpha^2)$ can be determined. One can verify this form easily using eqs.~(\ref{eq37}), (\ref{eq42}), and (\ref{eq43}). We define the “raising” and “lowering” operators as follows:
\begin{equation}
\label{eq45}
S^{op}_{\pm} = S_1^{op}\pm iS_2^{op};\;\;\bar{S}^{op}_{\pm} = \bar{S}_1^{op}\mp i\bar{S}_2^{op}.
\end{equation}
Note the change in signs for the rotating frame components. The spin eigenfunctions also satisfy
\begin{equation}
\label{eq46}
S^{op}_{\pm}\left|s,m_s,\bar{m}_s\right> = \hbar\sqrt{(s\mp m_s)(s\pm m_s + 1)}\left|s,m_s\pm 1,\bar{m}_s\right>,
\end{equation}
\begin{equation}
\label{eq47}
\bar{S}^{op}_{\pm}\left|s,m_s,\bar{m}_s\right> = \hbar\sqrt{(s\mp\bar{m}_s)(s\pm\bar{m}_s + 1)}\left|s,m_s,\bar{m}_s\pm 1\right>,
\end{equation}
except for phase factors that multiply the square roots but can be set equal to unity with no loss of generality. The relations~(\ref{eq42}), (\ref{eq43}), (\ref{eq44}), (\ref{eq46}), and (\ref{eq47}) may all be derived without reference to Euler angles from the commutation relations~(\ref{eq38}) and (\ref{eq40}) and the physical requirement that the maximum values of $|m_s|$ and $|\bar{m}_s|$ not exceed $s$; see any modern textbook on quantum mechanics, {\it e.g.} that by Shankar~\cite{Shankar}, for the conventional derivation that does not consider the rotating frame contributions. We chose the juxtaposition of signs in the second term of eq.~(\ref{eq45}) in order that $\bar{S}_+^{op}$ indeed acts to {\it raise} the index $\bar{m}_s$ by unity, etc. 

The statistical/Schr\"odinger wave equation for the system is
\begin{equation}
\label{eq48}
i\hbar\partial_t\psi = H^{op}\psi,
\end{equation}
where $H^{op}$ is given by eq.~(\ref{eq36}). At first glance, it might seem that the general solution could be written as a superposition of the eigenfunctions $\left|s,m_s,\bar{m}_s\right>$ over all allowed values of $s$, both integer and odd half-integer. However, that would violate the boundary condition mentioned above, that all {\it observable} functions of the azimuthal Euler angles must be single-valued in intervals of $2\pi$, {\it i.e.}, they must be periodic functions with period $2\pi$, despite the fact that the angles themselves have infinite range. A simple example suffices: Consider a superposition $\psi = C + D\exp{i\alpha^1/2}$, where $C$ and $D$ are nonzero functions of $x, t$, and the other angles, which is a superposition of $m_s = 0$ and $m_s = 1/2$ terms.  Then the “observable” probability density $\psi^*\psi$ is periodic in $\alpha^1$ with period $4\pi$, not $2\pi$. Reasoning from this example, it is easy to see that in order to ensure azimuthal angle periodicities of $2\pi$ for any probability density and for all other observable functions (which are always bilinear in $\psi^*$ and $\psi$), the general solution of eq.~(\ref{eq48}) must be written as a superposition of the spin eigenfunctions  with integer $s$ only, or with odd half-integer $s$ only. (Also, the integer-$s$ eigenfunctions are not orthogonal to the odd half-integer ones in azimuthal angle inervals $(0,2\pi)$, which reinforces the above restriction.) Since in this paper we are most interested in the spin-1/2 example, we adopt the superposition of the spin eigenfunctions with odd half-integer $s$ as the relevant solution of eq.~(\ref{eq48}). Furthermore, since the $H^{op}$ for a spherical model particle, eq.~(\ref{eq36}), contains the $S_i^{op}$ but not the $\bar{S}_i^{op}$, the sum over $\bar{m}_s$ is redundant; we may choose any allowed value of $\bar{m}_s$ with no loss of generality. Therefore, the relevant general solution of the statistical wave equation~(\ref{eq48}) may be written
\begin{equation}
\label{eq49}
\psi(x,\alpha,t) = \sum_{s=1/2,3/2,...}\sum_{m_s=-s}^s \psi^s_{m_s,\bar{m}_s}(x,t)\left|s,m_s,\bar{m}_s\right>
\end{equation}
where any odd-half-integer value of $\bar{m}_s,\;-s\le\bar{m}_s\le s$, may be used. Substitution of eq.~(\ref{eq49}) in eq.~({eq48}) yields $2s + 1$ coupled equations for each $s$.  In the next subsection, we write out the equations for $s = 1/2$.

     Before proceeding, we should discuss briefly why the spin eigenfunctions, which have the general form given by eq.~(\ref{eq44}), are themselves not required to be single-valued in azimuthal angle intervals $(0,2\pi)$, in contrast to eigenfunctions like $\exp{im\phi}$ involving the spherical polar or cylindrical coordinate azimuthal angle $\phi$. This question has been discussed often; a thoughtful treatment was given by Merzbacher~\cite{Merzbacher}. We paraphrase his answers as follows: The angle $\phi$ helps locate a point in Euclidean 3-space, so values of $\phi$ outside $(0,2\pi)$ are meaningless: $\phi$ and $\phi + 2n\pi$,   with $n$ an integer, are the same points, since e.g. they yield the same Cartesian coordinates for a given choice of the other spherical polar coordinates $(r,\theta)$. Therefore, the eigenfunctions themselves must satisfy periodic boundary conditions in the azimuthal angle interval $(0,2\pi)$, whereby $m$ must be an integer. (There is some difficulty with this answer in regard to the Aharonov-Bohm effect, which Merzbacher discusses). In contrast, for the azimuthal Euler angles, $\alpha^1$ and $\alpha^1 +2n\pi$ are not the same points, nor are $\alpha^3$ and $\alpha^3 + 2n\pi$, because the axis of rotation and/or the object may just keep on rotating, as mentioned above. (Merzbacher points out that the group space of the rotation group is a doubly connected space, whereby the the operation ``rotation by $2\pi$ about an axis" cannot be continuously deformed into the operation ``no rotation at all".)   Therefore, in the functions $\exp{im_s\alpha^1}$ and $\exp{i\bar{m}_s\alpha^3}$, the values of $(m_s,\bar{m}_s)$ are restricted to integers or odd half-integers only by the demands of the spin commutation rules, as discussed above, or, equivalently, by the requirement that all ``observable" functions be periodic in azimuthal Euler angle intervals $(0,2\pi)$.

     There is another consideration pertinent to the discussion just above. Consider the 2-dimensional subspace of the azimuthal Euler angles. One may always make the change of variables $\xi^1 = \mbox{$\frac{1}{2}$}(\alpha^1 + \alpha^3),\,\xi^3 = \mbox{$\frac{1}{2}$}(\alpha^1 - \alpha^3)$, {\it e.g.}, see the treatment of SU(2) by Arfken~\cite{Arfken}. The coordinates $(\xi^1, \xi^2=\alpha^2, \xi^3)$  provide a realization of the Cayley-Klein parameters. Then, for the fundamental intervals $0\le(\alpha^1,\alpha^3)\le 2\pi$, one obtains $0\le\xi^1\le 2\pi, -\pi\le\xi^3\le\pi$. Furthermore, the eigenfunctions of eq.~(\ref{eq44}) now involve $\exp{i[(m_s + \bar{m}_s)\xi^1 + (m_s - \bar{m}_s)\xi^3]}$, which {\it are} single-valued in their fundamental coordinate intervals because as discussed above $m_s$ and $\bar{m}_s$ must both be odd-half-integers, or both integers.  This coordinate transformation does not change any of the spin eigenvalues, and it also diagonalizes the metric in the spin space. 

     Also, it should again be noted that $H^{op}$ of eq.~(\ref{eq36}) contains the additional terms $(S^{op})^2/2I$ and $I(\tilde{g}B)^2/2$ compared to the conventional Hamiltonian operator. Consider the zero-momentum translational state of a “free” particle, i.e., of $H^{op}$ with the electromagnetic potentials equal to zero. Then $H^{op}$ is simply the rigid rotator Hamiltonian, which has energy eigenvalues $E_s = s(s+1)\hbar^2/2I$, whereby the energy required to produce a transition from $s=1/2$ to $s=3/2$ is $\Delta E = 3\hbar^2/2I$. Let $I=ma^2$, where $a$ is the approximate linear extension of the model rotator. For a nucleon, with $m\approx 10^3 MeV/c^2$ and $a\approx 1$ fm, $\Delta E\approx 50 MeV$, so (unstable) spin-3/2 baryons should exist, and they do. However, for an electron, with $m\approx 0.5 MeV/c^2$ and $a\le 10^{-2} fm$, $\Delta E\ge 10^9 MeV$. Unless a relativistic treatment can reduce this result by many orders of magnitude, one must conclude that creating a spin-3/2 lepton having the same extremely small linear extension as an electron is virtually impossible. Furthermore, any model of a charged object having semi-definite charge density, mass and intrinsic magnetic moment of the order of electronic values, and relevant linear extension $a$ smaller than about a fermi, rotating rigidly with angular speed $\omega$, and having spin angular momentum of order $\hbar$, seems to predict a linear surface speed $\omega a\gg c$~\cite{Griffiths}.
For this reason alone, many physicists feel that such models of particles must be discarded and spin regarded as {\it defined} by the commutation rules~eq.(\ref{eq38}). Nevertheless, it seems fascinating that the nonrelativistic Euler angle rigid rotator model plus conservation of probability actually predicts spin operators that do obey these commutation relations, and that there is no dilemma for $a\ge 1 fm$ and particles having nucleon mass or greater. Perhaps a relativistic extended particle model of leptons could resolve the dilemma, but detailed investigation of this possibility is beyond the scope of this work. However, in Appendix A we use another simple model of a spinning particle to show that the correct relativistic definition of the kinetic spin angular momentum yields $\omega a < c$  even for leptonic values of mass and particle size, while still predicting canonical spin angular momentum of order $\hbar$.

\subsection{Wave equation for spin 1/2}
\label{spin1/2}

We examine the solutions of eq.~(\ref{eq48}) for a given spin angular momentum. For this example, $s=1/2$, we may write
\begin{equation}
\label{eq50}
\psi^s_{\bar{m}_s}(x,\alpha,t) = U_+(x,t)\left|s,+,\bar{m}_s\right> + U_-(x,t)\left|s,-,\bar{m}_s\right>,
\end{equation}
for either allowed value of $\bar{m}_s$, as a suitable general solution of eq.~(\ref{eq48}). For notational convenience here, $\left|s,+,+\right>$ stands for $\left|\mbox{$\frac{1}{2}$},\mbox{$\frac{1}{2}$},\bar{m}_s\right>$, etc. The orthonormal spin eigenfunctions are proportional to the elements of the 2X2 irreducible representation of SU(2) in terms of the Euler angles~\cite{Arfken}:
\begin{equation}
\label{eq51}
\begin{array}{ccccc}
\left|s,+,+\right> & = & (2\pi)^{-1}\exp{[i(\alpha^1+\alpha^3)/2]}\cos{(\alpha^2/2)} & \equiv & u_{++}(\alpha); \\
\left|s,-,+\right> & = & (2\pi)^{-1}\exp{[i(-\alpha^1+\alpha^3)/2]}\sin{(\alpha^2/2)} & \equiv & u_{-+}(\alpha); \\
\left|s,+,-\right> & = & (2\pi)^{-1}\exp{[i(\alpha^1-\alpha^3)/2]}\sin{(\alpha^2/2)} & \equiv & u_{+-}(\alpha); \\
\left|s,-,-\right> & = & (2\pi)^{-1}\exp{[-i(\alpha^1+\alpha^3)/2]}\cos{(\alpha^2/2)} & \equiv & u_{--}(\alpha).
\end{array}
\end{equation}
It is very simple to verify that these eigenfunctions are indeed orthonormal under integration over the spin space (rotation group) volume, {\it e.g.}, that 
$$\left<s,+,+|s,+,+\right> = \int_0^{2\pi} d\alpha^1\int_0^{2\pi} d\alpha^3\int_0^\pi d\alpha^2\sin{\alpha^2}\,u^*_{++}u_{++} = 1,$$
etc. Using eqs.~(\ref{eq37}), (\ref{eq39}), and (\ref{eq41}), it is also straightforward to verify that these functions satisfy eqs.~(\ref{eq42}), (\ref{eq43}), (\ref{eq46}), and (\ref{eq47}) for $s=1/2$.

We write the Hamiltonian operator~(\ref{eq36}) as 
\begin{equation}
\label{eq52}
H^{op} = H_1^{op} + H_2^{op},
\end{equation}
where
\begin{equation}
\label{eq53}
H_1^{op} = \mbox{$\frac{1}{2m}$}(\bm{p}^{op}-q\bm{A}/c)^2 +q\varphi + \mbox{$\frac{1}{2I}$}(\bm{S}^{op})^2 + \mbox{$\frac{1}{2}$}I\tilde{g}^2 B^2;
\end{equation}
\begin{equation}
\label{eq54}
H_2^{op} = -\tilde{g}\bm{B}\bm{\cdot}\bm{S}^{op} = -\tilde{g}[B_3S_3^{op} + \mbox{$\frac{1}{2}$}(B_-S_+^{op}+B_+S_-^{op})] 
\end{equation}
Here, $S_{\pm}^{op}$ are given by eq.~(\ref{eq45}), and $B_{\pm} = B_1\pm iB_2$. Since we may use either value of $\bar{m}_s$, we choose $+1/2$. Then a little algebra, using eqs.~(\ref{eq42}), (\ref{eq43}), (\ref{eq46}), (\ref{eq47}), (\ref{eq48}), and (\ref{eq50})-(\ref{eq52}), yields two coupled equations,
\begin{equation}
\label{eq55}
\begin{array}{ccc}
(-i\hbar\partial_t + H_1^{op})U_+ - \mbox{$\frac{1}{2}$}\hbar\tilde{g}(B_3U_+ + B_-U_-) & = & 0; \\
(-i\hbar\partial_t + H_1^{op})U_- + \mbox{$\frac{1}{2}$}\hbar\tilde{g}(B_3U_- - B_+U_+) & = & 0.
\end{array}
\end{equation}
Of course, these two coupled equations may be written as a 2X2 matrix equation. If one defines the matrices
\begin{equation}
\label{eq56}
\sigma_1 =\left(\begin{array}{cc} 0 & 1 \\ 1 & 0 \end{array}\right);\;\;\;\sigma_2 =\left(\begin{array}{cc} 0 & -i \\ i & 0 \end{array}\right);\;\;\;\sigma_3 =\left(\begin{array}{cc} 1 & 0 \\ 0 & -1 \end{array}\right),
\end{equation}
which happen to be the Pauli spin matrices, and also defines the column matrix (spinor)
\begin{equation}
\label{eq57}
(U) = \left(\begin{array}{c} U_+ \\ U_- \end{array}\right),
\end{equation}
then one obtains the matrix equation
\begin{equation}
\label{eq58}
(-i\hbar\partial_t + H_1^{op})(U) - (\tilde{g}B_i)(\mbox{$\frac{1}{2}$}\hbar\sigma_i)(U) = 0.
\end{equation}
This is the Pauli-Schrödinger equation, except that $H_1^{op}$ contains the additional terms $\mbox{$\frac{1}{2I}$}(\bm{S}^{op})^2 + \mbox{$\frac{1}{2}$}I\tilde{g}^2 B^2$, which must be present for the reasons discussed above. Also, this equation is valid for any value of $g$ in the magnetogyric ratio $\tilde{g}=gq/2mc$, not just for the value $g = 2$ originally chosen for the electron in order to match atomic spectral data. Note that as usual the definition $\tilde{S}_i^{op}=\mbox{$\frac{1}{2}$}\hbar\sigma_i$ yields the conventional matrix representation of the space-fixed Cartesian components $S_i$ of the spin angular momentum operator. Also, we remark again that exactly the same eqs.~(\ref{eq55}) and (\ref{eq58}) result if we use the spin eigenfunctions for $\bar{m}_s=-1/2$  instead of those for $+1/2$ . It is also noteworthy that the “magic” factorization $(\bm{p}^{op}-q\bm{A}/c)^2\rightarrow[\bm{\sigma}\bm{\cdot}(\bm{p}^{op}-q\bm{A}/c)]^2$ of the translational Hamiltonian (see {\it e.g.} Sakurai~\cite{Sakurai}), the nonrelativistic analog of the Dirac factorization, allows only $g = 2$ and also does not provide the terms $\mbox{$\frac{1}{2I}$}(\bm{S}^{op})^2 + \mbox{$\frac{1}{2}$}I\tilde{g}^2 B^2$ . Since $g = 2$ is correct for the electron without radiative corrections, and since that factor arises from the factorization of the translational Hamiltonian, perhaps the electron spin actually originates from translational zitterbewegung induced by the SZPF, as has been proposed~\cite{BarutBr, BarutZ}. After all, as the material in the appendices implies, it is not really clear whether angular momentum is “orbital” or “spin” in nature. However, see the discussion in Sec.~\ref{summadis} below.  

\section{Summary and discussion}
\label{summadis}

\subsection{Summary}
\label{summary}

This work concerned a system of one particle having mass, charge, spin angular momentum, and associated magnetic moment, which for nonrelativistic motions requires a six-dimensional metric space for three CM coordinates and three Euler angle coordinates. In Sec. II we followed the development by R. Young~\cite{Young} and showed that the magnetic field appears in the classical nonrelativistic Hamiltonian as a gauge field associated with the space-fixed frame Cartesian components of the spin angular momentum conjugate to the Euler angles. In section III, we applied the general rules derived in section III of [I] to obtain the six-dimensional Schrödinger equation from the classical Hamiltonian. We found that the space-fixed Cartesian components of the canonical spin angular momentum become operators that are linear combinations of derivatives with respect to the Euler angles and obey the conventional commutation rules for quantum angular momentum. We also found that the body-fixed Cartesian components obey “left-handed” commutation rules. Furthermore, we showed that the particle may access odd half-integer or integer spin eigenstates, but not both. This result follows from applying physical boundary conditions in the Euler angle description; it doesn’t seem to follow from the spin commutation relations alone. For a particular spin index $s$, we showed that the general six-dimensional Schrödinger equation yields $2s+1$ coupled equations for the amplitudes $psi^s_{m_s,\bar{m}_s}(x,t)$  in eq.~(\ref{eq49}). We also showed that, for spin 1/2, these coupled equations reduce to the Pauli-Schrödinger equation with the usual Pauli matrix representation of the space-fixed system spin operators, but with arbitrary magnetogyric ratio and additional rigid rotator terms in the Hamiltonian. In Appendix~\ref{appA}, we provided a simple example of a rigidly rotating charged particle for which the kinetic spin angular momentum is treated relativistically. This treatment showed that the tangential linear speeds associated with the rotation are always subluminal, no matter how small the mass or effective radius of the model particle. In Appendix~\ref{appB}, we showed that the usual spin commutation rules apply to a rigid rotator of any structure, not just to a spherically symmetric rotator, which implies that any object in a rigid rotator eigenmode must have spin equal to one of the eigenvalues $s = (0,1/2,1,3/2,...)$, regardless of its internal structure or how many subparticles it contains, etc. However, the rotational Hamiltonian and its eigenfunctions and eigenvalues do depend on the particle structure. 

\subsection{Discussion}
\label{discussion}

One topic that seems to merit discussion is the distinction between spin and orbital angular momentum. The distinction is fuzzy at best. For example, consider the classical model of an extended object as a cloud containing many point particles; see {\it e.g.} Goldstein's textbook~\cite{Goldstein}. The total angular momentum of the cloud may always be defined with respect to an arbitrarily chosen origin of coordinates, and it can always be written (nonrelativistically) as the orbital angular momentum of the total mass, located at the CM, as it moves about the origin, plus the sum of the orbital angular momenta of each of the constituent point particles about the CM. So in general all the angular momentum in such a model is orbital. The total orbital angular momentum about the CM is called the “intrinsic” angular momentum of the object. If the point particles in the cloud are rigidly bound to each other, so that all their motions relative to the CM can be represented in terms of a single angular velocity vector that requires only three Euler angle coordinates to describe, then that intrinsic orbital angular momentum is called the “spin” angular momentum. Thus, a distinction between orbital and spin angular momentum doesn't really exist for this “Goldstein model” of a rigid body: It's all orbital! 

     Given the above, how can we justify allowing half-odd-integer spin angular momentum, but only integer orbital angular momentum? One response, adopted by many, is that we cannot justify it, so we’ll simply go with the predictions of the commutation rules, to be used as needed. Another possible response is to note that the Goldstein model may not be the appropriate one. In actuality, all collections of particles and thus all rigid bodies are comprised of electrons, quarks, (and photons and gluons and…), as far as we know. The fermions and bosons in this mix have their own irreducible “spins”, analogously to subvortices within larger vortices in fluid mechanics, not the case for the subparticles in the Goldstein model. It would seem that only in a model utilizing fundamental subparticles having irreducible spins can one hope to distinguish between orbital and spin angular momentum. So this argument turns into an argument for the existence of fundamental particles with irreducible spins (not that we really need that argument!). (In relativistic field theory, the distinction between orbital and spin angular momentum follows from Noether’s theorem; see e.g. the paper by the author on stress-energy tensors~\cite{Goedecke}. It has been suggested that the irreducible spins of the fundamental particles may originate in their orbital zitterbewegung driven by the SZPF~\cite{BarutBr, BarutZ}. This idea might be worth pursuing further.

\appendix

\section{Relativistic rotation}
\label{appA}

As mentioned above, one objection to a nonrelativistic extended spinning electron model having semidefinite charge density is that it seems to require a tangential linear speed $v$ at its outer boundary that far exceeds $c$. In this appendix we treat such a model relativistically and show that then $v < c$ for all parameter choices. Since the calculation is for illustration only, we choose an extremely simple model, namely, a circular ring of radius $a$ with mass $m$ and charge $q$ uniformly distributed around the ring. Let the ring lie in the $x-y$ plane and be constrained to rotate about the $z$-axis, its symmetry axis, with angular velocity $z$-component $\omega_z = \dot{\alpha}$, where $\alpha (t)$ is the relevant Euler angle. If one neglects the self magnetic field and there are no applied fields, the nonrelativistic expression for the $z$-component of the canonical spin angular momentum is simply $S_z = ma^2\dot{\alpha}$. If this expression must have magnitude $\hbar/2$, one obtains $\beta = |\dot{\alpha}|a/c = \hbar/2mac$. For electron mass $m\approx 10^{-27}gm$ and radius $a\le 10^{-2} fm$, one obtains $\beta\ge 10^4$, which reveals the source of the objection. 

     As a general rule, if one assumes a speed to be slow, but then solves for it and finds supraluminal values, one should start over using relativistic expressions. For this model, it is clear that any infinitesimal ring segment of mass $dm$ has tangential linear momentum $\gamma a \dot{\alpha}dm$, where $\gamma = (1 - a^2\dot{\alpha}^2/c^2)^{-1/2}$. Therefore, in the absence of magnetic fields the spin angular momentum of the ring is $S_z = m\gamma a^2 \dot{\alpha}$. For $|S_z| = \hbar/2$, one obtains $\beta\gamma = \hbar/2mac \equiv \Lambda$, whereby $\beta = \Lambda/\sqrt{1 + \Lambda^2} < 1$.  For the electron parameter values above, $\Lambda\ge 10^4$, whereby $\beta \approx 1 - 0.5\times 10^{-8}$.  For baryon parameter values, $m\ge 1.8\times 10^{-24} gm$, $a\ge 1 fm$, one obtains $\beta\le 0.1$, which is essentially nonrelativistic. 

In this simple ring model it seems that the angular momentum may be regarded as either spin or orbital, which recalls the question of whether half-odd-integer spin is allowed. Note that to model the rotating ring properly one must include all three Euler angles in the classical description, which allows odd-half-integer spin as discussed above. Also, a fully relativistic treatment, well beyond the scope of this work, is needed to identify both translational and rotational motions and their contributions to the effective magnetic moments of electrons and other leptons. 

\section{Arbitrary rigid rotator}
\label{appB}

In this appendix we first show that the spin operator commutation rules for a rigidly rotating object of arbitrary structure are the same as for a spherically symmetric object.  Then we obtain the Hamiltonian of a charged rotating body of arbitrary structure interacting with a magnetic field.

\subsection{Spin commutation rules}
\label{commrules}

We make use of the fact that one may express the inertia tensor of any object in a body-fixed principal axis Cartesian frame in which the tensor is diagonal with principal moments of inertia $I_i, i=1,2,3$. The rotational kinetic energy is 
\begin{equation}
\label{eqB1}
T_{rot} = \mbox{$\frac{1}{2}$}\bar{I}_i\bar{\omega}_i^2 = \mbox{$\frac{1}{2}$}mg_{bc}\dot{\alpha}^b\dot{\alpha}^c,
\end{equation}
where the second term extends the summation convention to indices repeated twice, which is notationally convenient. Applying eq.~(\ref{eq4}) of the text yields
\begin{equation}
\label{eqB2}
g_{bc} = m^{-1}\bar{I}_i b_{ib}b_{ic}.
\end{equation}
Here, $m$ is a parameter having dimension mass that may be chosen as the mass of the object, $g_{bc}$ is the covariant metric in the 3-space of the Euler angles $\alpha^b$, and the $b_{ib}$ are given by eq.~(\ref{eq5}). It is easy to see that the covariant and contravariant basis vectors in the Euler angle space are given by
\begin{equation}
\label{eqB3}
\bm{e}_b = (\bar{I}_i/m)^{1/2}b_{ib}\hat{\bar{\bm{e}}}_i;\;\;\;\bm{e}^b = (m/\bar{I}_i)^{1/2}b_{bi}^{-1}\hat{\bar{\bm{e}}}_i,
\end{equation}
where the $\hat{\bar{\bm{e}}}_i$ are the Cartesian unit basis vectors in the body-fixed frame. These relations are the analogs of eqs.~(\ref{eq21}) and (\ref{eq22}). Then
\begin{equation}
\label{eqB4}
g^{bc} = \bm{e}^b\bm{\cdot}\bm{e}^c = m(\bar{I}_i)^{-1} b^{-1}_{bi}b^{-1}_{ci}.
\end{equation}
It is also easy to see that $\bm{e}_b\bm{\cdot}\bm{e}^c = \delta_b^c$ and that $g^{bc}g_{ca} = \delta_a^b$, as must be the case. Now consider a freely rotating particle, so that its rotational Lagrangian $L_{rot} = T_{rot}$. Than the conjugate (angular) momenta are $P_b = \partial L_{rot}/\partial\dot{\alpha}^b = mg_{bc}\dot{\alpha}^c$, whereby
\begin{equation}
\label{eqB5}
\dot{\alpha}^b = m^{-1}g^{bc}P_b.
\end{equation}
Using eqs.~(\ref{eq4}), (\ref{eqB4}), and (\ref{eqB5}), one obtains
\begin{equation}
\label{eqB6}
\bar{S}_1 = \bar{I}_1\bar{\omega}_1 = b^{-1}_{c1}P_c,
\end{equation}
and similarly for the other spin angular momentum components in the body-fixed frame. Then, with $P_c\rightarrow p^{op}_c = -i\hbar\partial/\partial\alpha^c$ as in eq.~(\ref{eq35}), one obtains
\begin{equation}
\label{eqB7}
\bar{S}_i\rightarrow\bar{S}^{op}_i = -i\hbar b^{-1}_{ci}\partial/\partial\alpha^c.
\end{equation}
These operators are identical to those defined by eq.~(\ref{eq39}), which means that they are independent of the structure of the rigidly rotating particle. Furthermore, since $b_{ib} = R_{ij}a_{jb}$ and $\bar{S}^{op}_i = R_{ij}S^{op}_j$, one also obtains 
\begin{equation}
\label{eqB8}
S^{op}_i = -i\hbar a^{-1}_{ci}\partial/\partial\alpha^c,
\end{equation}
the same as eq.~(\ref{eq37}). Therefore, the commutation relations of eqs.~(\ref{eq38}) and (\ref{eq40}) are still valid, and the simultaneous eigenfunctions of $(\bm{S}^{op})^2, S^{op}_3, \bar{S}^{op}_3$  are still the $\left|s,m_s,\bar{m}_s\right>$ that satisfy eqs.~(\ref{eq42}), (\ref{eq43}), (\ref{eq46}), and (\ref{eq47}). However, the free particle Hamiltonian operator is easily seen to be
\begin{equation}
\label{eqB9}
H^{op}_{rot} = \mbox{$\frac{1}{2}$}\bar{I}_i^{-1}\bar{S}^{op}_i\bar{S}^{op}_i.
\end{equation}
If all three principal moments of inertia are different, the individual spin eigenfunctions are not eigenfunctions of $H^{op}_{rot}$. However, suppose the object has a rotational symmetry axis, which we may always choose to be the body-fixed $z$-axis, so that $\bar{I}_1 = \bar{I}_2 \ne \bar{I}_3$. Then
\begin{equation}
\label{eqB10}
H^{op}_{rot} = \mbox{$\frac{1}{2}$}[\bar{I}_1^{-1}\bar{S}^{op}_i\bar{S}^{op}_i + (\bar{I}_3^{-1} - \bar{I}_1^{-1})\bar{S}^{op}_3\bar{S}^{op}_3].
\end{equation}
The individual spin eigenfunctions $\left|s,m_s,\bar{m}_s\right>$ are eigenfunctions of this Hamiltonian; the eigenvalues are
\begin{equation}
\label{eqB11}
E_{s,\bar{m}_s} = \mbox{$\frac{1}{2}$}[\bar{I}_1^{-1}s(s+1)\hbar^2 + (\bar{I}_3^{-1} - \bar{I}_1^{-1})\bar{m}_s^2\hbar^2].
\end{equation}
This result reveals that some of the degeneracy of the spherical rigid rotator energy eigenvalues for a given $s$ may be removed for an axially but not spherically symmetric object, with the energy shifts dependent on the quantum number associated with the body-fixed frame. 

We emphasize again that the spin operators, their commutation relations, their eigenfunctions and eigenvalues, and the raising and lowering operator relations~(\ref{eq47}) and (\ref{eq48}) are independent of the principal moments of inertia and any other model particle parameters. This is not a surprising result in view of the fact that the rotation of any rigid object is described in terms of three angles that simply relate the time-dependent orientation of a rotating orthonormal triad of basis vectors relative to a nonrotating triad. This fact implies that any object must have half-odd-integer or integer spin when it is in a rigid rotation mode, regardless of how many subparticles it contains and how they are distributed, which does seem to be the case for baryons and nucleons. For example, the three quarks in a nucleon are now thought to have both spin angular momentum and orbital angular momentum about the nucleon center of momentum, and the gluons may contribute to the total angular momentum as well, but all that internal structure must arrange itself so that the total spin of a nucleon in its ground state is 1/2.  

\subsection{Interaction with a magnetic field}
\label{magfield}

Consider a rigidly rotating charged extended particle of arbitrary structure, except that we consider here only cases for which the center of charge is the same point as the CM. The Cartesian components of the particle’s intrinsic magnetic moment in the rotating system are expressible as linear functions of the rotating system angular velocity components:
\begin{equation}
\label{eqB12}
\bar{\mu}_i = \bar{Q}_{ij}\bar{\omega}_j,  
\end{equation}
where $\bar{Q}_{ij} = \bar{Q}_{ji}$ are constants, components of a symmetric rank two tensor that could be obtained from the definition of eq.~(\ref{eq8}) for the magnetic moment of a given current density $\bm{J}$ that is not spherically symmetric about its center, instead of eq.~(\ref{eq9}). Then for a particle with its CM position $\bm{X}$ at rest, the relevant Lagrangian is
\begin{equation}
\label{eqB13}
L_{rot} = \mbox{$\frac{1}{2}$}\bar{I}_i\bar{\omega}_i\bar{\omega}_i + \bar{Q}_{ij}\bar{\omega}_i\bar{B}_j,
\end{equation}
where $\bar{B}_j = R_{jk}B_k$ are defined just after eq.~(\ref{eq24}); they are effective rotating frame Cartesian components of any magnetic field that may be present. These components are evaluated at $\bm{X}$, and they also depend on the Euler angles. The spin angular momentum conjugate to $\bar{\omega}_i$ is
\begin{equation}
\label{eqB14}
\bar{S}_i = \partial L_{rot}/\partial\bar{\omega}_i = \bar{I}_i\bar{\omega}_i + \bar{Q}_{ij}\bar{B}_j,
\end{equation}
which from the definition $H_{rot} = \bar{\omega}_i\bar{S}_i - L_{rot}$ yields the rotational Hamiltonian
\begin{equation}
\label{eqB15}
H_{rot} = \mbox{$\frac{1}{2}$}\bar{I}_i^{-1}(\bar{S}_i - \bar{Q}_{ik}\bar{B}_k)(\bar{S}_i - \bar{Q}_{il}\bar{B}_l).
\end{equation}
The apparent interaction Hamiltonian, which is the cross-term in the quadratic form above, is
\begin{equation}
\label{eqB16}
H_{int} = -\mbox{$\frac{1}{2}$}\bar{I}_i^{-1}\bar{Q}_{ik}(\bar{B}_k)(\bar{S}_i + \bar{S}_i\bar{B}_k).
\end{equation}
Clearly, this interaction is a more complicated form than in the spherically symmetric case, and we cannot go further than eq.~(\ref{eqB16}) in the case of arbitrary particle structure. For illustration, consider the special case in which $\bar{Q}_{ik} = \delta_{ik}\bar{Q}_i$, so that the prinipal axis coordinate system for the charge distribution is the same as that for the mass distribution. Also, let the ratio
\begin{equation}
\label{eqB17}
\bar{Q}_i/\bar{I}_i = \tilde{g}
\end{equation}
be the same for each principal axis, where $\tilde{g} = gq/2mc$ as in eq.(\ref{eq14}). Then 
\begin{equation}
\label{eqB18}
H_{int} = -\mbox{$\frac{1}{2}$}\tilde{g}(\bar{B}_i\bar{S}_i + \bar{S}_i\bar{B}_i).
\end{equation}
When $\bar{S}_i$ is replaced by $\bar{S}_i^{op}$, it is not algebraically trivial to show that the commutator $[\bar{S}_i^{op},\bar{B}_i] = 0$. The proof reduces to showing that $R_{ij}(S_j^{op}R_{jk}) = 0$, which can be done using eq.~(\ref{eqB8}) and several equations in the text, or by starting with the definition of the 3-vector angular velocity, eq.~(\ref{eq3}). Therefore, in this quite special but still not spherically symmetric case, the effective interaction Hamiltonian is the same as for a spherically symmetric particle, namely, $H_{int} = -\tilde{g}\bar{B}_i\bar{S}_i = -\tilde{g}B_i S_i$.


\begin{thebibliography}{99}

\bibitem{Bopp} F. Bopp and R. Haag, Z. Naturforsch. 5a, 644 (1950).
\bibitem{Nyborg} P. Nyborg, Nuovo Cimento 23, 47 (1962).
\bibitem{Corben} H. C. Corben, ``Classical and Quantum Theories of Spinning Particles" (Holden-Day, 
      San Francisco, 1968), and many references therein.
\bibitem{delaPena} L. de la Pe\~{n}a-Auerbach, J. Math. Phys. 12, 453 (1971).
\bibitem{Young} R. Young, Am. J. Phys. 44, 581 (1976).
\bibitem{BarutBr} A. Barut and A. Bracken, Phys. Rev. D23, 2454 (1981), and D24, 3333 (1981).
\bibitem{Jauregi} R. J\'{a}uregi and L. de la Pe\~{n}a, Phys. Lett. A86, 280 (1981).
\bibitem{BarutZ} A. Barut and N. Zanghi, Phys. Rev. Lett. 52, 2009 (1984).
\bibitem{BarutBoMa} A. Barut, M. Bo\v{z}i\'{c}, and Z. Mari\'{c}, Ann. Phys. 214, 53 (1992).
\bibitem{Arsenovic} D. Arsenovic, A. Barut, Z. Mari\'{c}, and M. Bo\v{z}i\'{c}, Nuovo Cimento B 110, 163 (1995).
\bibitem{Arfken} G. Arfken, ``Mathematical Methods for Physicists", 3rd ed. (Academic Press, NY, 1985).
\bibitem{MathewsW} J. Mathews and R. Walker, ``Mathematical Methods of Physics", 2nd ed. (W. A. 
      Benjamin, NY, 1970).
\bibitem{Shankar} R. Shankar, ``Principles of Quantum Mechanics", 2nd ed. (Springer, NY, 1994). 
\bibitem{Merzbacher} E. Merzbacher, Am. J. Phys. 30, 237 (1962).
\bibitem{Griffiths} D. Griffiths, ``Introduction to Quantum Mechanics", 2nd ed. (Prentice-Hall, London,
      2005).
\bibitem{Sakurai} J. Sakurai, ``Advanced Quantum Mechanics" (Addison-Wesley, Reading, MA, 1967).
\bibitem{Goldstein} H. Goldstein, ``Classical Mechanics", 2nd ed. (Addison-Wesley, Reading, MA, 1980).
\bibitem{Goedecke} G. Goedecke, J. Math. Phys. 15, 72, 1974.


\end{thebibliography}
\end{document}